\documentstyle[12pt]{article}
\begin{document}
\def\bea{\begin{eqnarray}}
\def\eea{\end{eqnarray}}
\title{\bf {
 The Cardy-Verlinde formula and entropy of  black holes in de
Sitter spaces }}
\author{
M.R. Setare  \footnote{E-mail: rezakord@ipm.ir}
  \\
 {Institute for Theoretical Physics and
Mathematics, Tehran, Iran}}
%\date{\small{\today}}
\maketitle
\begin{abstract}
In this paper we show that the entropy of a cosmological horizon
in topological Reissner-Nordstr\"om- de Sitter and Kerr-Newman-de
Sitter spaces can be described by the Cardy-Verlinde formula,
which is supposed to be an entropy formula of conformal field
theory in any number of dimension. Furthermore, we find that the
entropy of a black hole horizon can also be rewritten in terms of
the Cardy-Verlinde formula for these black holes in de Sitter
spaces, if we use the definition due to Abbott and Deser for
conserved charges in asymptotically de Sitter spaces. Such result
presume a well-defined dS/CFT correspondence, which has not yet
attained the credibility of its AdS analogue.

 \end{abstract}
% \begin{document}
\newpage
% \vspace*{10mm}

 \section{Introduction}
The holographic duality which connects $n+1$-dimensional gravity
in Anti-de Sitter (AdS) background with $n$-dimensional conformal
field theory (CFT) has been discussed vigorously for some
years\cite{AdS}. But it seems that we live in a universe with a
positive cosmological constant which will look like de Sitter
space--time in the far future. Therefore, we should try to
understand quantum gravity or string theory in de Sitter space
preferably in a holographic way. Of course, physics in de Sitter
space is interesting even without its connection to the real
world; de Sitter entropy and temperature have always been
mysterious aspects of quantum gravity\cite{GH}.\\
While string theory successfully has addressed the problem of
entropy for black holes, dS entropy remains a mystery. One reason
is that the finite entropy seems to suggest that the Hilbert space
of quantum gravity for asymptotically de Sitter space is finite
dimensional, \cite{{Banks:2000fe},{Witten:2001kn}}.
 Another, related, reason is that the horizon and entropy in
de Sitter space have an obvious observer dependence. For a black
hole in flat space (or even in AdS) we can take the point of view
of an outside observer who can assign a unique entropy to the
black hole. The problem of \ what an observer venturing inside the
black hole experiences, is much more tricky and has not been given
a satisfactory answer within string theory. While the idea of
black hole complementarity provides useful clues, \cite
{Susskind}, rigorous calculations are still limited to the
perspective of the outside observer. In de Sitter space there is
no way to escape the problem of the observer dependent entropy.
This contributes to the difficulty of de Sitter space.\\
More recently, it has been proposed that defined in a manner
analogous to the AdS/CFT correspondence,  quantum gravity in a de
Sitter (dS) space is dual to a certain
 Euclidean  CFT living on a spacelike boundary of the
dS space~\cite{Strom} (see also earlier works
\cite{Hull}-\cite{Bala}). Following this proposal, some
investigations on the dS space have been carried out
recently~\cite{Mazu}-\cite{Ogus}. According to the dS/CFT
correspondence, it might be expected that as in the case of AdS
black holes~\cite{Witten2}, the thermodynamics of cosmological
horizon in asymptotically dS spaces can be identified with that
of a certain Euclidean CFT residing on a spacelike boundary of the
asymptotically dS spaces.\\
One of the remarkable outcomes of the AdS/CFT and dS/CFT
correspondence has been the generalization of Cardy's formula
(Cardy-Verlinde formula) for arbitrary dimensionality, as well as
a variety of AdS and dS backgrounds. In this paper, we will show
that the entropy of a cosmological horizon in the topological
Reissner-Nordstr\"om -de Sitter (TRNdS) and topological
Kerr-Newman-de Sitter spaces (TKNdS) can also be rewritten in the
form of the Cardy-Verlinde formula.  We then show that if one uses
the Abbott and Deser (AD) prescription \cite{AD}, the entropy of
black hole horizons in dS spaces can also be expressed by the
Cardy-Verlinde formula.

\section{Topological Reissner-Nordstr\"om-de Sitter Black Holes}
We start with an $(n+2)$-dimensional TRNdS black hole solution,
whose metric is
\begin{eqnarray}
&& ds^2 = -f(r) dt^2 +f(r)^{-1}dr^2 +r^2 \gamma_{ij}dx^{i}dx^{j}, \nonumber \\
&&~~~~~~ f(r)=k -\frac{\omega_n M}{r^{n-1}} +\frac{n \omega_n^2
Q^2}{8(n-1) r^{2n-2}}
     -\frac{r^2}{l^2},
\end{eqnarray}
where
\begin{equation}
\omega_n=\frac{16\pi G_{n+2}}{n\mbox {Vol}(\Sigma)},
\end{equation}
where $\gamma_{ij}$ denotes the line element of an $n-$dimensional
hypersurface $\Sigma$ with constant curvature $n(n-1)k$ and volume
$Vol(\Sigma)$ , $G_{n+2}$ is the $(n+2)-$dimensional Newtonian
gravity constant, $M$ is an integration constant, $Q$ is the
electric/magnetic charge of Maxwell field. When $k=1$, the metric
Eq.(1) is just the Reissner-Nordstr\"om-de Sitter solution. For
general $M$ and $Q$, the equation $f(r)=0$ may have four real
roots. Three of them are real, the largest on is the cosmological
horizon $r_{c}$, the smallest is the inner (Cauchy) horizon of
black hole, the one in between is the outer horizon $r_{+}$ of the
black hole. And the fourth is negative and has no physical
meaning. The case $M=Q=0$ reduces
to the de Sitter space with a cosmological horizon $r_{c}=l$.\\
When $k=0$ or $k<0$, there is only one positive real root of
$f(r)$, and this locates the position of cosmological horizon
$r_{c}$.\\
In the case of $k=0$, $\gamma_{ij}dx^{i}dx^{j}$ is an
$n-$dimensional Ricci flat hypersurface, when $M=Q=0$ the solution
Eq.(1) goes to pure de Sitter space
\begin{equation}
ds^{2}=\frac{r^{2}}{l^{2}}dt^{2}-\frac{l^{2}}{r^{2}}dr^{2}+r^{2}dx_{n}^{2}
\end{equation}
in which $r$ becomes a timelike coordinate.\\
When $Q=0$, and $M\rightarrow -M$ the metric Eq.(1)is the TdS
(topological de Sitter) solution \cite{{cai1},{med}}, which have a
cosmological horizon and a naked singularity, for this type of
solution, the Cardy-Verlinde formula also work well.
\\
Here we review the BBM prescription \cite{BBM} for computing the
conserved quantities of asymptotically de Sitter spacetimes
briefly. In a theory of gravity, mass is a measure of how much a
metric deviates near infinity from its natural vacuum behavior;
i.e, mass measures the warping of space. Inspired by the analogous
reasoning in AdS space \cite{{by},{b}} one can construct a
divergence-free Euclidean quasilocal stress tensor in de Sitter
space by the response of the action to variation of the boundary
metric we have
\begin{eqnarray}
T^{\mu \nu} &=& {2 \over \sqrt{h}} { \delta I \over \delta h_{\mu
\nu}} = \ \  {1 \over 8\pi G} \left[ K^{\mu\nu} - K \, h^{\mu\nu}
+ {n \over l} \, h^{\mu\nu} +\frac{l}{n}  \,G^{\mu\nu} \right] ,
 \label{stressminus}
\end{eqnarray}
where $h^{\mu\nu}$ is the metric induced on surfaces of fixed
time, $K_{\mu\nu}$, $K$ are respectively extrinsic curvature and
its trace, $G^{\mu\nu}$ is the Einstein tensor of the boundary
geometry. To compute the mass and other conserved quantities, one
can write the metric $h^{\mu\nu}$ in the following form
\begin{equation}
    h_{\mu\nu} \, dx^{\mu} \, dx^{\nu } =
       N_{\rho}^{2} \, d\rho^{2} +
       \sigma_{ab}\, (d\phi^a + N_\Sigma^a \, d\rho) \,
               (d\phi^b + N_\Sigma^b \, d\rho)
%            N_{\phi}^{2} \, (d\phi + V \,
%       d\rho)^{2} \, .
       \label{boundmet}
\end{equation}
where the $\phi^{a}$ are angular variables parametrizing closed
surfaces around the origin. When there is a Killing vector field
$\xi^{\mu}$ on the boundary, then the conserved charge associated
to $\xi^{\mu}$ can be written as \cite{{by},{b}}
\begin{equation}
   Q =  \oint_{\Sigma}  d^{n}\phi \,\sqrt{\sigma } \,
   n^{\mu}\xi^{\mu} \,T_{\mu\nu}
   \label{chargedef}
\end{equation}
where $n^{\mu}$ is the unit normal vector on the boundary,
$\sigma$ is the determinant of the metric $\sigma_{ab}$. Therefore
the mass of an asymptotically de Sitter space is as
\begin{equation}
    M =
    \oint_{\Sigma}  d^{n}\phi \,\sqrt{ \sigma } \, N_{\rho} \,
\epsilon
    ~~~~~;~~~~~ \epsilon \equiv
    n^{\mu}n^{\nu} \,
    T_{\mu\nu} \, ,
    \label{massdef}
\end{equation}
where Killing vector is normalized as $\xi^{\mu} = N_{\rho}
n^{\mu}$. Using this prescription \cite{BBM}, the gravitational
mass, having subtracted the anomalous Casimir energy, of the TRNdS
solution is
\begin{equation}
\label{3eq3} E=-M =-\frac{r_c^{n-1}}{\omega_n} \left (k
-\frac{r_c^2}{l^2} +
    \frac{n\omega_n^2 Q^2}{8(n-1)r_c^{2n-2}}\right).
\end{equation}
 Some thermodynamic
quantities associated with the cosmological horizon are
\begin{eqnarray}
 && T_{c}= \frac{1}{4\pi r_c} \left(-(n-1)k +(n+1)\frac{r_c^2}{l^2}
    +\frac{n\omega_n^2 Q^2}{8 r_c^{2n-2}}\right), \nonumber \\
&& S_{c} =\frac{r_c^n\mbox{Vol}(\sigma)}{4G}, \nonumber \\
&& \phi_{c} =-\frac{n}{4(n-1)}\frac{\omega_n Q}{r_c^{n-1}},
\end{eqnarray}
where $\phi_{c}$ is the chemical potential conjugate to the charge
$Q$. \\
 The Casimir energy $E_C$,
defined as $E_C =(n+1) E-nTS-n\phi Q$ in this case, is found to be
\begin{equation}
E_C=-\frac{2nkr_c^{n-1}\mbox{Vol}(\sigma)}{16\pi G},
\end{equation}
when $k=0$, the Casimir energy vanishes, as the case of
asymptotically AdS space. When $k=\pm 1$, we see from Eq.(10) that
the sign of energy is just contrast to the case of TRNAdS space
\cite{youm}.\\
 Thus we can see that the entropy
 Eq.(9)of the cosmological horizon can be rewritten as
 \begin{equation}
 S=\frac{2\pi l}{n}\sqrt{|\frac{E_{C}}{k}|(2(E-E_q)-E_C)},
\end{equation}
where
\begin{equation}
E_q = \frac{1}{2}\phi_{c} Q =-\frac{n}{8(n-1)}\frac{\omega_n
Q^2}{r_c^{n-1}}.
\end{equation}
We note that  the entropy expression (11) has a similar form as
the
case of TRNAdS black holes \cite{youm}.\\
 For the black hole
horizon, which is only for the case $k=1$, associated
thermodynamic quantities are
\begin{eqnarray}
\label{3eq8} && T_{b}=\frac{1}{4\pi r_b}\left( (n-1)
-(n+1)\frac{r_{b}^2}{l^2} -\frac{n\omega_n^2 Q^2}
   {8r_{b}^{2n-2}}\right), \nonumber \\
&&  S_{b}=\frac{r_{b}^n \mbox{Vol}(\sigma)}{4G}, \nonumber \\
&&  \phi_{b} =\frac{n}{4(n-1)}\frac{\omega_n Q}{r_{b}^{n-1}}.
\end{eqnarray}
Now if we uses the BBM mass Eq.(\ref{3eq3}) the black hole horizon
entropy cannot be expressed in a form like Cardy-Verlinde formula
\cite{cai1}. The other way for computing conserved quantities of
asymptotically de Sitter space is the Abbott and Deser (AD)
prescription \cite{AD}. According to this prescription, the
gravitational mass of asymptotically de Sitter space coincides
with the ADM mass in asymptotically flat space, when the
cosmological constant goes to zero. Using the AD prescription for
calculating conserved quantities the black hole horizon entropy of
TKNdS space can be expressed in term of the Cardy-Verlind formula
\cite{cai1}. The AD mass of TRNdS solution can be expressed in
terms of black hole horizon radius $r_b$ and charge $Q$,
\begin{equation}
  E' =M =\frac{r_{b}^{n-1}}{\omega_n} \left
(1-\frac{r_{b}^2}{l^2} +
   \frac{n\omega_n^2 Q^2}{8(n-1)r_{b}^{2n-2}}\right).
\end{equation}
In this case, the Casimir energy, defined as
 $ E'_C
=(n+1) E' -n T_{b}
 S_{b}-n \phi_{b} Q$, is
\begin{equation}
  E'_C =\frac{2n r_{b}^{n-1}\mbox{Vol}(\sigma)}{16\pi
G},
\end{equation}
and the black hole entropy $ S_{b}$ can be rewritten as
\begin{equation}
\label{3eq11}  S_{b} =\frac{2\pi l}{n}\sqrt{ E'_C |2(
E'-E'_q)-E'_C|},
\end{equation}
where
\begin{equation}
 E'_q =\frac{1}{2} \phi_{b} Q=\frac{n\omega_n
Q^2}{8(n-1)r_{b}^{n-1}},
\end{equation}
which is the energy of electromagnetic field outside the black
hole horizon. Thus we demonstrate that the black hole horizon
entropy of TRNdS solution can be expressed in a form as the
Cardy-Verlinde formula. However, if one uses the BBM mass Eq.(8),
the black hole horizon entropy $S_b$ cannot be expressed by a form
like the Cardy-Verlinde formula.
\section{Topological Kerr-Newman-de Sitter Black Holes}
The line element of TKNdS black holes in 4-dimension case is given
by
\begin{eqnarray}
ds^{2} &=&-\frac{\Delta _{r}}{\rho ^{2}}\left(dt-\frac{a}{\Xi
}\sin ^{2}\theta d\phi \right)^{2}+\frac{\rho ^{2}}{\Delta
_{r}}dr^{2}+\frac{\rho ^{2}}{\Delta
_{\theta }}d\theta ^{2} \nonumber  \\
&&+\frac{\Delta _{\theta }\sin ^{2}\theta }{\rho ^{2}}\left[a
dt-\frac{ (r^{2}+a^{2})}{\Xi }d\phi \right]^{2},  \label{kdsmet}
\end{eqnarray}
where
\begin{eqnarray}
\Delta _{r} &=&(r^{2}+a^{2})\left(k-\frac{r^{2}}{l^{2}}\right)
-2Mr+q^2,  \nonumber \\
\Delta _{\theta } &=&1+\frac{a^{2}\cos ^{2}\theta}{ l^{2}},
  \nonumber \\
\Xi &=&1+\frac{a^{2}}{l^{2}},  \nonumber \\
\rho ^{2} &=&r^{2}+a^{2}\cos ^{2}\theta .  \label{kdelt}
\end{eqnarray}
Here the parameters $M$, $a$, and $q$ are associated with the
mass, angular momentum, and electric charge parameters of the
space-time, respectively. The topological metric Eq.(18) will only
solve the Einstein equations if k=1, which is the spherical
topology. In fact when $k=1$, the metric Eq.(18) is just the
Kerr-Newman -de Sitter solution. Three real roots of the equation
$\Delta _{r}=0$, are the locations of three horizons, the largest
being the cosmological horizon $r_{c}$, the smallest is the inner
horizon of black hole, the one in between is the outer
horizon $r_{b}$ of the black hole.\\
If we want in the $k=0,-1$ cases to solve the Einstein equations,
then we must set $sin\theta \rightarrow \theta$, and $sin\theta
\rightarrow sinh\theta$ respectively \cite{man1}-\cite{man4}. When
$k=0$ or $k=-1$, there is only one positive real root of $\Delta
_{r}$, and this locates the position of cosmological horizon
$r_{c}$.
\\
In the BBM prescription\cite{BBM}, the gravitational mass,
subtracted the anomalous Casimir energy, of the 4-dimensional
TKNdS solution is
\begin{equation}
E=\frac{-M}{\Xi}. \label{bbmass}
\end{equation}
Where the parameter $M$ can be obtained from  the equation
$\Delta_{r}=0$. On this basis, the following relation for the
gravitational mass can be obtained
\begin{equation}
E=\frac{-M}{\Xi} =\frac{(r_c^2+a^2)(r_c^2-k l^2)-q^2 l^2}{2\Xi r_c
l^2}. \label{kmass}
\end{equation}
The Hawking temperature of the cosmological horizon  is given by
\begin{equation}
T_{c}=\frac{-1}{4\pi}\frac{\Delta
'_{r}(r_{c})}{(r_c^2+a^2)}=\frac{3r_c^4+r_c^2(a^2-k
l^2)+(ka^2+q^2)l^2}{4\pi r_c l^2(r_c^2+a^2)}. \label{tem}
\end{equation}
The entropy associated with the cosmological horizon can be
calculated as
\begin{equation}
S_{c}=\frac{\pi(r_{c}^{2}+a^{2})}{\Xi}.
 \label{ent}
\end{equation}
The angular velocity of the cosmological horizon is given by
\begin{equation}
\Omega_{c}=\frac{-a\Xi}{(r_{c}^{2}+a^{2})}.
 \label{ang}
\end{equation}
The angular momentum $J_{c}$, the electric charge $Q$, and the
electric potentials $\phi_{qc}$ and $\phi_{qc0}$ are given by
\begin{eqnarray}
  & & {\mathcal{J}}_c=\frac{M a}{\Xi^2}, \nonumber \\
 & & Q =\frac{q}{\Xi}, \nonumber  \\
 & & \Phi_{qc}=-\frac{q r_c}{r_c^2+a^2}, \nonumber  \\
 & & \Phi_{qc0}=-\frac{q}{r_c},
 \label{ctherm}
\end{eqnarray}
The obtained above quantities of the cosmological horizon satisfy
the first law of thermodynamics:
\begin{equation}
dE=T_cdS_c+\Omega_c d{\mathcal{J}}_c+(\Phi_{qc}+\Phi_{qc0}) dQ .
\label{Flth}
\end{equation}

 Using the
Eqs.(\ref{ent},\ref{ctherm}) for the cosmological horizon entropy,
angular momentum and charge, and also the equation
$\Delta_{r}(r_{c})=0$, we can obtain the mtric parameters $M$,
$a$, $q$ as a function of $S_{c}$, ${\mathcal{J}}_c$  and $Q$, and
after that we can write $E$ as a function of these
thermodynamical quantities: $E(S_{c},{\mathcal{J}}_c,Q)$ (see
\cite{cal}). Then one can define the quantities conjugate to
$S_{c}$, ${\mathcal{J}}_c$ and $Q$, as
\begin{equation}
T_c=\left( \frac{\partial E}{\partial S_c}\right) _{J_c,_Q},\ \
\Omega_c =\left( \frac{\partial E}{\partial {J}_c}\right)
_{S_c,Q},\ \ \Phi_{qc} =
\left( \frac{\partial E%
}{\partial Q}\right) _{S_c,J_c}\\\Phi_{qc0} =lim_{a\rightarrow 0}
\left( \frac{\partial E%
}{\partial Q}\right) _{S_c,J_c}\\  \label{Dsmar}
\end{equation}
 Making use of the fact
that the metric for the boundary CFT can be determined only up to
a conformal factor, we rescale the boundary metric for the CFT to
the following form:
\begin{equation}
ds_{CFT}^2=\lim_{r \rightarrow \infty}\frac{R^2}{r^2}ds^2 ,
\label{euniverse}
\end{equation}
Then the thermodynamic relations between the boundary CFT and the
bulk TKNdS are given by
\begin{equation}
E_{CFT}=\frac{l}{R}E,\hspace{0.07 cm}T_{CFT}=\frac{l}{R}T
,\hspace{0.07 cm}J_{CFT}=\frac{l}{R}J,\hspace{0.07
cm}\phi_{CFT}=\frac{l}{R}\phi,\hspace{0.07
cm}\phi_{0CFT}=\frac{l}{R}\phi_{0}, \label{CFT}
\end{equation}
 The Casimir energy $E_C$,
defined as $E_C =(n+1)
E-n(T_cS_c+J_c\Omega_c+Q/2\phi_{qc}+Q/2\phi_{qc0})$ , and $n=2$ in
this case, is found to be
\begin{equation}
E_C=-\frac{k(r_c^2+a^2)l }{R \Xi r_c}, \label{ckecas01}
\end{equation}
in KNdS space case \cite{jing} the Casimir energy $E_c$ is always
negative, but in TKNdS space case the Casimir energy can be
positive, negative or vanishing depending on the choice of $k$.
 Thus we can see that the entropy
 Eq.(\ref{ent})of the cosmological horizon can be rewritten as
 \begin{equation}
 S=\frac{2\pi R}{n}\sqrt{|\frac{E_{c}}{k}|(2(E-E_q)-E_c)},\label{careq}
\end{equation}
where
\begin{equation}
E_q = \frac{1}{2}\phi_{c0} Q.\label{qeq}
\end{equation}
We note that  the entropy expression (\ref{careq}) has a similar
form as in the
case of TRNdS black holes Eq.(11).\\
 For the black hole
horizon, which  only exists for the case $k=1$ the associated
thermodynamic quantities are
\begin{equation}
T_{b}=\frac{1}{4\pi}\frac{\Delta
'_{r}(r_{b})}{(r_b^2+a^2)}=-\frac{3r_b^4+r_b^2(a^2-l^2)+(a^2+q^2)l^2}{4\pi
r_b l^2(r_b^2+a^2)}. \label{tem2}
\end{equation}
\begin{equation}
S_{b}=\frac{\pi(r_{b}^{2}+a^{2})}{\Xi}.
 \label{ent2}
\end{equation}
\begin{equation}
\Omega_{b}=\frac{a\Xi}{(r_{b}^{2}+a^{2})}.
 \label{ang2}
\end{equation}
\begin{equation}
   {\mathcal{J}}_b=\frac{M a}{\Xi^2},
\end{equation}
\begin{equation}
Q =\frac{q}{\Xi},
\end{equation}
\begin{equation}
\Phi_{qb}=\frac{q r_b}{r_b^2+a^2},
 \end{equation}
\begin{equation}
\Phi_{qb0}=\frac{q}{r_b}.
\end{equation}
 The AD mass of TKNdS solution can be expressed in
terms of the black hole horizon radius $r_b$, $a$ and charge $q$:
\begin{equation}
  E'=\frac{M}{\Xi} =\frac{(r_b^2+a^2)(r_b^2-l^2)-q^2 l^2}{2\Xi r_b l^2}.
\label{kmass2}
\end{equation}
The quantities obtained above  of the black hole horizon also
satisfy the first law of thermodynamics:
\begin{equation}
dE'=T_bdS_b+\Omega_b d{\mathcal{J}}_b+(\Phi_{qb}+\Phi_{qb0}) dQ .
\label{Flth1}
\end{equation}

The thermodynamics quantities of the CFT must be rescaled by a
factor $\frac{l}{R}$ similar to the pervious case. In this case,
the Casimir energy, defined by
 $ E'_C
=(n+1) E' -n(T_bS_b+J_b\Omega_b+Q/2\phi_{qb}+Q/2\phi_{qb0})$, is
\begin{equation}
 E'_C =\frac{(r_b^2+a^2)l }{R \Xi r_b}, \label{ckecas02}
\end{equation}
and the black hole entropy $ S_{b}$ can be rewritten as
\begin{equation}
S_{b}=\frac{2\pi
R}{n}\sqrt{E'_{C}|(2(E'-E'_q)-E'_C)|},\label{careq2}
\end{equation}
where
\begin{equation}
E'_q =\frac{1}{2}\phi_{qb0} Q.\label{qeq2}
\end{equation}
This is the energy of an electromagnetic field outside the black
hole horizon. Thus we demonstrate that the black hole horizon
entropy of the TKNdS solution can be expressed in the form of the
Cardy-Verlinde formula. However, if one uses the BBM mass
Eq.(\ref{kmass}) the black hole horizon entropy $S_{b}$ cannot be
expressed in a form like the Cardy-Verlinde formula. Our result is
in favour of the dS/CFT correspondence.
\section{Conclusion}
The Cardy-Verlinde formula recently proposed by  Verlinde
\cite{Verl}, relates the entropy of a  certain CFT to its total
energy and Casimir energy in arbitrary dimensions. In the spirit
of dS/CFT correspondence, this formula has been shown to hold
exactly for the cases of dS Schwarzschild, dS topological, dS
Reissner-Nordstr\"om , dS Kerr, and dS Kerr-Newman  black holes.
In this paper we have further checked the Cardy-Verlinde formula
with topological Reissner-Nordstr\"om- de Sitter and
topological Kerr-Newman  de Sitter black holes. \\
It is well-known that there is no black hole solution whose event
horizon is not a sphere, in a de Sitter background, although there
are such solutions in an anti-de Sitter background; then in TRNdS,
TKNdS spaces for the case k=0,-1 the black hole does not have an
event horizon, however the cosmological horizon geometry is
spherical, flat and hyperbolic for k=1,0,-1, respectively. As we
have shown there exist two different temperatures and entropies
associated with the cosmological horizon and black hole horizon,
in TRNdS, TKNdS spacetimes. If the temperatures of the black hole
and cosmological horizon are equal, then the entropy of the system
is the sum of the entropies of cosmological and black hole
horizons. The geometric features of the black hole temperature and
entropy seem to imply that the black hole thermodynamics is
closely related to nontrivial topological structure of spacetime.
In \cite{cai2} Cai, et al in order to relate the entropy with the
Euler characteristic $\chi$ of the corresponding Euclidean
manifolds have presented the following relation:
\begin{equation}
S=\frac{\chi_{1}A_{BH}}{8}+\frac{\chi_{2}A_{CH}}{8},
\end{equation}
in which the Euler number of the manifolds is divided into two
parts; the first part comes from the black hole horizon and the
second part come from the cosmological horizon (see also
\cite{{teit},{gib},{lib}}). If one uses the BBM mass of the
asymptotically dS spaces, the black hole horizon entropy cannot be
expressed in a form like the Cardy-Verlinde formula\cite{cai1}. In
this paper, we have found that if one uses the AD prescription to
calculate conserved charges of asymptotically dS spaces, the
black hole horizon entropy can also be rewritten in the form of
the Cardy-Verlinde formula, which is indicates that the
thermodynamics of the black hole horizon in dS spaces can also be
described by a certain CFT. Our result is also reminiscent of
Carlip's claim~\cite{Carlip}(to see a new formulation which is
free of the inconsistencies encountered in Carlip's in.\cite{mu})
that for  black holes of any dimensionality the Bekenstein-Hawking
entropy can be reproduced using the Cardy formula~\cite{Cardy}.
Also we have shown that the Casimir energy for a cosmological
horizon in TKNdS space case can be positive, negative or
vanishing, depending on the choice of $k$; by contrast, the
Casimir energy for a cosmological horizon in KNdS space is always
negative \cite{jing}.

\end{document}